\documentclass[aps,twocolumn]{revtex4-2}
\usepackage[dvipdfmx]{graphicx}
\usepackage[dvipdfmx]{color}
\usepackage{dcolumn}
\usepackage{bm}
\usepackage{amsmath}

\newcommand{\Figref}[1]{Fig.~\ref{#1}}
\newcommand{\Figsref}[1]{Figs.~\ref{#1}}
\newcommand{\Figureref}[1]{Figure~\ref{#1}}
\newcommand{\Eqref}[1]{Eq.~(\ref{#1})}

\newcommand{\Eqsref}[1]{Eqs.~(\ref{#1})}
\newcommand{\Braref}[1]{(\ref{#1})}
\newcommand{\Appref}[1]{Appendix~\ref{#1}}

\newcommand{\Secref}[1]{Sec.~\ref{#1}}

\begin{document}


\title{Modeling of chemically active particles at an air-liquid interface}


\author{Shun Imamura${}^{1,2}$}
\email{imamura@cmpt.phys.tohoku.ac.jp}
\author{Toshihiro Kawakatsu${}^{1}$}
\affiliation{${}^{1}$Department of Physics, Graduate School of Science, Tohoku University, Sendai 980-8578, Japan}
\affiliation{${}^{2}$Mathematics for Advanced Materials-OIL, AIST-Tohoku University, Sendai 980-8577, Japan}


\date{\today}

\begin{abstract}
The collective motion of chemically active particles at an air-liquid interface is studied theoretically as a dynamic self-organization problem.
Based on a physical consideration, we propose a minimal model for self-propelled particles by combining hydrodynamic interaction, capillary interaction, driving force by Marangoni effect, and Marangoni flow.
Our model has successfully captured the features of chemically active particles, that represent dynamic self-organized states such as crystalline, chain, liquid-like and spreading states.
\end{abstract}


\maketitle

\section{Introduction}
The collective behavior of chemically active particles in fluids is important not only for the development of active matter physics\cite{vicsek2012collective, marchetti2013hydrodynamics, gompper20202020, popescu2020chemically}, which focuses on the collective phenomena of self-propelled particles, but also for a deeper understanding of the dynamic self-organization of systems that move by chemical mechanisms such as bacteria\cite{lauga2009hydrodynamics, koch2011collective}.
Chemically active particles have been realized using colloids\cite{paxton2004catalytic, ebbens2010pursuit, bechinger2016active, zottl2016emergent}, camphor particles\cite{kohira2001synchronized, nagayama2004theoretical, kitahata2004self, nakata2015physicochemical, boniface2019self}, and droplets\cite{hanczyc2007fatty, toyota2009self, tanaka2015spontaneous, maass2016swimming}.

Experimentally, the collective behavior of chemically active particles, such as camphor particle system\cite{soh2008dynamic, ikura2013collective, nishimori2017collective} and droplet system\cite{nagai2005mode, chen2009self, tanaka2017dynamic}, constrained on an air-liquid interface are particularly interesting as models of two-dimensional wet active matter systems\cite{marchetti2013hydrodynamics}.
The wet system is defined as those systems that are mainly dominated by hydrodynamic interaction.
Here, hydrodynamic interaction has a long-range nature and a lack of action-reaction law, and can be an important factor in creating complex collective behaviors\cite{kano2017mathmatical}.
For example, in an experiment by Soh {\it et al.}, camphor particles placed in a circular container show a dynamic self-organization where the particles are arranged with a constant interval when the population of particles is large\cite{soh2008dynamic}.
Tanaka {\it et al.} observed complex dynamics in droplet systems, where the behavior of the droplets changes with time, as if they dance spontaneously\cite{tanaka2015spontaneous, tanaka2017dynamic}.
The physical mechanisms of these complex collective dynamics are still unknown.

The motion of the chemically active particles at an air-liquid interface was explained based on the following characteristic physical mechanisms\cite{kohira2001synchronized, soh2008dynamic, yabunaka2012self, ikura2012suppression, masoud2014collective, nishi2015bifurcation, nakata2015marangoni, matsuda2016acceleration, dominguez2016collective, yabunaka2016collision, kitahata2018effective, hirose2020two};
\begin{itemize}
\item  hydrodynamic interaction between particles (HI)\cite{rotne1969variational, yamakawa1970transport, perkins1991hydrodynamic},
\item  lateral capillary force (capillary interaction) due to the deformation of the air-liquid interface (CF)\cite{dominguez2016collective},
\item  self-propelling force induced by the inhomogeneity in the surface tension due to the gradient of the surrounding surfactant concentration field (SP)\cite{kohira2001synchronized, yabunaka2012self, boniface2019self},
\item  interaction between particles mediated by the concentration field of surfactant (IC)\cite{kohira2001synchronized},
\item  Marangoni flow inside the liquid induced by the change in the surface tension of the air-liquid interface due to surfactants (MF)\cite{ikura2012suppression, nakata2015marangoni, dominguez2016effective, masoud2014collective, matsuda2016acceleration, kitahata2018effective},
\end{itemize}
where the last 3 effects are originating from the concentration gradient of the surfactant, and hereafter we will refer them as ``Marangoni effects''.

In a colloidal particle system, Masoud and Shelley dealt with HI, IC, and MF using direct numerical simulation\cite{masoud2014collective}, and Dominguez {\it et al.} dealt with HI, CF, and MF analytically using the reaction diffusion equation\cite{dominguez2016collective}.
For the camphor particle system, Soh {\it et al.} dealt with HI, SP, IC, and MF numerically\cite{soh2008dynamic}, and Hirose {\it et al.} solved the diffusion equations for CF, SP, and IC numerically to analyze the particle motion\cite{hirose2020two}.
For self-propelled droplet systems, Yabunaka and Yoshinaga studied the motion of two-particle systems by numerically and analytically treating concentrations and fluid flow field, by incorporating HI, SP, and IC\cite{yabunaka2016collision}.
Despite these successes in describing the collective behaviors, the coupling between the concentration field and the fluid flow field makes it difficult to identify the main factors of the physical phenomena.
It is also difficult to perform the direct numerical calculations of many-particle systems.
Therefore, it is important to incorporate each physical element into a simple coarse-grained model in a way that the elements can be turned on and off analytically. 

In the present study, we propose a minimal model that can deal with interacting many self-propelled particles while incorporating all elementary physical mechanisms (hydrodynamic interaction, capillary interaction, and Marangoni effect).
The advantage of our model is that we need to solve neither the concentration field nor the fluid flow fields, which reduces the computational cost considerably.

The present article is organized as follows. 
First, we show the detail of our proposed model in \Secref{sec: model}.
Then, the simulation method is explained in \Secref{sec:simulation method}.
In \Secref{sec:Simulation results}, we report the results of the simulations on single-particle, two-particle and many-particle systems.
Finally, in \Secref{sec:conclusion}, we draw our conclusion.

\section{Model}
\label{sec: model}
\begin{figure}[htbp]
  \centering
  \includegraphics[clip, width=8.0cm]{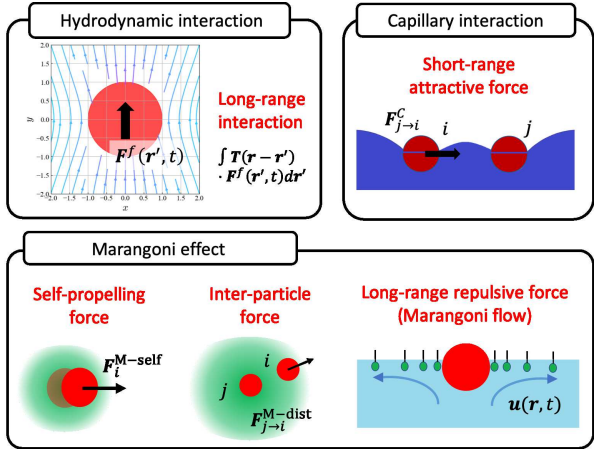}
  \caption{
  Five physical mechanisms introduced in our model;
  long-range hydrodynamic interaction (RPY type mobility tensor with mirror image method applied),
  short-range capillary interaction,
  self-propelling force induced by the Marangoni effect, 
  short-range interaction due to the concentration field of surfactant, and
  Marangoni flow that produces long-range repulsive force.
    }
  \label{fig:intro_model}
\end{figure}

Here, we construct a model of self-propelled particles by introducing the hydrodynamic interaction, the capillary interaction, and the Marangoni effects.
For the Marangoni effect, we treat the self-propelling force due to the gradient of the surfactant concentration field and the fluid flow due to the difference of the interfacial tension of the air-liquid interface, separately.
\Figureref{fig:intro_model} shows schematic illustrations of individual physical processes.

To describe the equations of motion for the particles, surfactant concentration, and the liquid flow, we introduce a cartesian coordinate system where the $z$-axis is set to the vertical direction to the air-liquid interface and $x$ and $y$ axes in the interface (i.e. $z=0$).
We also assume that the air region and the liquid region correspond to $z > 0$ and $z < 0$, respectively.

The equations of motion for the $i$-th particle are given by
\begin{align}
m \frac{d}{dt} \bm{v}_{i}^{p}(t) &= \bm{F}_{i}(t) + \bm{K}_{i}(t),
\label{equ:equation of motion}
\\
\frac{d}{dt} \bm{r}_{i}^{p}(t) &= \bm{v}_{i}^{p}(t),
\end{align}
where $m$ is mass of a particle and $\bm{r}_{i}^{p}(t)$ and $\bm{v}_{i}^{p}(t)$ represent the position and the velocity of $i$-th particle at time $t$, respectively.
$\bm{F}_{i}$ and $\bm{K}_{i}$ represent the driving force and the viscous drag force, respectively.
We model the Marangoni effects based on an assumption that surfactant is emitted and diffuses from inside the particle.
To describe such a diffusion process, we assume a reaction-diffusion equation for the surfactant emitted from an isolated particle located at the origin of the system as follows\cite{nakata2000spontaneous, hayashima2001camphor};
\begin{equation}
\frac{\partial c}{\partial t}
+
\bm{\nabla} \cdot \left( c \bm{v}^{f} \right)
=
D \bm{\nabla}^{2} c
- \kappa \left( c - c_{\infty} \right)
+ A \delta(r)
,
\label{equ:reaction-diffusion eq.}
\end{equation}
where, $c(\bm{r}, t)$ represents the concentration field of the surfactant, and $c_{\infty}$ is its value at the infinite distance from the particle in the steady state.
In \Eqref{equ:reaction-diffusion eq.}, the first, second and third terms on the right-hand side represent diffusion, consumption, and emission of the surfactant, respectively, where $D, \kappa$ and $A$ are positive constants that represent the diffusion coefficient, the consumption rate of surfactant that dissolves into the bulk water from its surface per unit time, and the emission rate of the surfactant.
In this model of the surfactant diffusion, we approximate the particle as a point particle for the sake of the analytical convenience\cite{nagayama2004theoretical}.

By solving the reaction-diffusion equation, \Eqref{equ:reaction-diffusion eq.}, in the steady state,  we can obtain the following concentration field,
\begin{align}
c(\bm{r})
=
\frac{A}{2 \pi D} \exp{\left[ \frac{\bm{r} \cdot \bm{v}^{f}}{2D} \right]}
K_{0}\left(2 r \sqrt{\frac{\kappa}{4D} + \left( \frac{\bm{v}^{f}}{4D} \right)^{2}} \right)
,
\label{equ:concentration field for delta function}
\end{align}
where $K_{0}(x)$ is the modified Bessel function of the second kind of order 0.
For the derivation of \Eqref{equ:concentration field for delta function}, see \Appref{sec: The derivation of Green function for the concentration field of surfactant}.
The concentration field \Eqref{equ:concentration field for delta function} decays exponentially with the characteristic diffusion length $\lambda = \sqrt{D/\kappa}$.
We should note that, by assuming the source of the surfactant as the delta function form, the solution of \Eqref{equ:reaction-diffusion eq.} corresponds to the Green’s function, with which we can obtain the time evolution of the surfactant concentration field emitted from a source with any shape, such as a step function or a Gaussian function, by using the convolution integration.

We introduce the driving force $\bm{F}_{i}$ for $i$-th particle as
\begin{equation}
\bm{F}_{i}(t)
=
\bm{F}_{i}^{\text{M}}(t)
+
\sum_{j (j \neq i)}
\left(
\bm{F}_{j \to i}^{\text{C}}(t)
+
\bm{F}_{j \to i}^{\text{exc}}(t)
\right),
\label{equ:driving force}
\end{equation}
where $\bm{F}^{\text{M}}_{i}$ represents the force acting on the $i$-th particle caused by the Marangoni effect, and $\bm{F}^{\text{C}}_{j \to i}(t)$ and $\bm{F}^{\text{exc}}_{j \to i}(t)$ are the lateral capillary force and the excluded volume effect from $j$-th particle to $i$-th particle, respectively. 
In the following, we will describe the detail of each force separately.

The Marangoni force $\bm{F}^{\text{M}}$ is divided into two parts as $\bm{F}^{\text{M}}_{i} = \bm{F}^{\text{M-self}}_{i} + \bm{F}^{\text{M-dist}}_{i}$ (See \Appref{sec: The driving force caused by Marangoni effect} for details of calculation).
$\bm{F}^{\text{M-self}}_{i}$ is the self-propelling force given by
\begin{align}
\bm{F}^{\text{M-self}}_{i}
= - \frac{\alpha A R}{D}
I_{1} \left( \frac{v^{f}_{i} R}{2D} \right)
K_{0} \left( W_{i} R \right)
\hat{\bm{v}}^{f}_{i}
,
\label{equ:self-propelling force}
\end{align}
where $v^{f}_{i}$ is the fluid flow velocity at the position of $i$-th particle, $R$ is the radius of the particle, $\alpha$ is the reduction rate interfacial tension due to surfactant defined in \Eqref{equ:the surface tension for dependence on surfactant concentration}, $W_{i} = 2 \sqrt{\frac{\kappa}{4D} + \left( \frac{v^{f}_{i}}{4D} \right)^2}$, and $I_{1}(x)$ is the modified Bessel function of the first kind of order 1.
We should note that, when the particle is moving with a velocity $\bm{v}^{p}_{i}(t)$, the replacement of $\bm{v}^{f}_{i}$ by $\bm{v}^{f}_{i} - \bm{v}^{p}_{i}$ should be made.
Therefore, in \Eqref{equ:self-propelling force}, we should regard $\bm{v}^{f}_{i} = \bm{v}^{f}(\bm{r}^{p}_{i}(t), t)$.
On the other hand, $\bm{F}^{\text{M-dist}}_{i} = \sum_{j \neq i} \bm{F}^{\text{M-dist}}_{j \to i}$ is the interaction force due to the surfactant emitted by the neighboring particles given by
\begin{align}
 \bm{F}^{\text{M-dist}}_{j \to i}
 &=
 - \frac{\alpha A R^{2}}{2D}
 \exp{\left[ \frac{\bm{v}^{f}_{i} \cdot \bm{r}_{ij}}{2D}\right]}
 \nonumber
 \\
 &\ 
 \times \left[
 K_{0}(W_{j} r_{ij}) \frac{\bm{v}^{f}_{i}}{2D}
 -
 W_{j} K_{1}(W_{j} r_{ij}) \hat{\bm{r}}_{ij}
 \right].
 \label{equ: concentration field interaction}
\end{align}
where $\bm{r}_{i j}(t) = \bm{r}^{p}_{i}(t) - \bm{r}^{p}_{j}(t)$, $r_{ij} = \vert \bm{r}_{ij} \vert$ and $\hat{\bm{r}}_{i j} = \bm{r}_{i j}/ r_{ij}$ is the unit vector in the relative direction between $i$-th and $j$-th particles, and $K_{1}(x)$ is the modified Bessel function of the second kind of order 1.

$\bm{F}^{\text{C}}_{j \to i}(t)$ in the second term of \Eqref{equ:driving force} represents the driving force caused by the lateral capillary force\cite{kralchevsky1994capillary} from $j$-th particle to $i$-th particle given by
\begin{equation}
 \bm{F}^{\text{C}}_{j \to i}(t) = - 2 \pi \gamma_{0} q Q_{i} Q_{j} K_{1}(q r_{ij}) \hat{\bm{r}}_{i j}(t) \ \ \ (i \neq j),
 \label{equ:lateral capillary force}
\end{equation}
where $\gamma_{0}$ represents the surface tension of the bare air-liquid interface without surfactant, and $q^{-1}$ the capillary length and $Q_{i}$ the so-called ``capillary charge'' of the $i$-th particle, defined by $Q_{i} = R^{c}_{i} \sin{\psi_{i}}$.
$R^{c}$ represents the radius of contact line between three phases, i.e. air-liquid-particle, and $\psi$ represents the constant angle at the meniscus between the particle and the air-liquid interface.

Finally, $\bm{F}^{\text{exc}}_{j \to i}(t)$ in the second term of \Eqref{equ:driving force} represents the excluded volume effect given by the WCA potential\cite{weeks1971role} between contacting particles.

We assume an overdamped dynamics where the driving force $\bm{F}_{i}(t)$ in \Eqref{equ:driving force} is balanced by the viscous drag force $\bm{K}_{i}$ defined by
\begin{align}
\bm{K}_{i}(t)
&=
- \zeta
\left(
\bm{v}^{p}_{i}(t)
-
\bm{v}^{f}_{i}(t)
\right),
\\
\bm{v}^{f}_{i}(t)
&=
\bm{u}(\bm{r}^{p}_{i}(t))
+
\int d \bm{r}' \bm{T}(\bm{r}^{p}_{i} - \bm{r}') \cdot \bm{F}^{f}(\bm{r}', t)
\label{equ: velocity field}
,
\\
\bm{F}^{f}(\bm{r}, t)
&=
\sum_{i} \zeta \left( \bm{v}^{p}_{i}(t) - \bm{v}^{f}_{i}(t) \right) \delta( \bm{r}^{p}_{i}(t) - \bm{r} )
\label{equ: force field acting on the fluid}
,
\end{align}
where $\zeta$ is the friction coefficient of the particle floating on the liquid-air interface.
We assume that the fluid velocity at the $i-$th particle position $\bm{v}^{f}_{i}(t)$ is described by the Stokes equation $\mu \bm{\nabla}^{2} \bm{v}^{f}(\bm{r}, t) = \bm{\nabla} P(\bm{r}, t) - \bm{F}^{f}(\bm{r}, t)$, where the inertia term is neglected and the incompressibility condition $\bm{\nabla} \cdot \bm{v}^{f}(\bm{r}, t) = 0$ is imposed.
Here, $\mu$, $P(\bm{r}, t)$ and $\bm{F}^{f}(\bm{r}, t)$ represent the viscosity of the fluid, the pressure field and the external force field, respectively.
$\bm{u}(\bm{r}, t)$ denotes the Marangoni flow. (See \Appref{sec:Repulsive interaction caused by the Marangoni flow} for details of calculation.)
$\bm{T}(\bm{r})$ is the Green's function for hydrodynamic interaction under the boundary condition that the fluid velocity in the $z$-direction vanishes at the air-liquid interface ($z=0$) and is given by
$\bm{T}(\bm{r} - \bm{r}')
= \bm{G}(\bm{r} - \bm{r}')
+ \bm{G}(\bm{r} - \bm{r}' + 2h \bm{e}_{z}) \cdot \bm{P}_{z},$
where $h$ is the $z$-coordinate of the center of mass of the particle,
$\bm{P}_{z} = \bm{1} - 2 \bm{e}_{z} \bm{e}_{z}$ represents the reflection operator by a mirror image\cite{perkins1991hydrodynamic} and $\bm{e}_{z}$ is the unit vector in the $z$-direction.
$\bm{G}(\bm{r})$ expresses the RPY type mobility tensor\cite{rotne1969variational, yamakawa1970transport}
\begin{align}
	\bm{G}(\bm{r})
	=
	\frac{1}{8 \pi \mu r}
	\left[
	( \bm{1} + \hat{\bm{r}}\hat{\bm{r}} )
	+
	\frac{1}{3}\left( \frac{R}{r} \right)^{2}
	( \bm{1} - 3 \hat{\bm{r}}\hat{\bm{r}} )
	\right],
\end{align}
where the first term on the right-hand side is a stokeslet, and the second term is a source doublet.
The boundary condition at the air-fluid interface is specified by the balance between the forces acting on the interface as
$ \left. \frac{\partial v^{f}_{x}}{\partial z} \right|_{z = 0} =
\left. \frac{\partial v^{f}_{y}}{\partial z} \right|_{z = 0} = 0,
\ v^{f}_{z}|_{z = 0} = 0$.

\section{Simulation method}
\label{sec:simulation method}
In the present study, we consider the case that the center of mass of each particle is located on the air-liquid interface, i.e., $h \sim 0$, where the effective velocity field at $xy$-plane is described by $\bm{T}(\bm{r}) \sim 2 \bm{G}(\bm{r})$.
Due to such a simplification, the parameter $\zeta$, which is given by  $\zeta \sim 3 \pi \mu R$, is the friction constant of the particle at the air-liquid interface\cite{dorr2016drag}.
Although our model is a minimal model on the coarse-grained scale, it is still too complicate to be solved analytically. 
Thus, we introduce a further simplification that $R_{c}$ and $\psi$ are common to all the particles, i.e., $R^{c}_{i} = R$ and $\psi_{i} = \psi$.
Therefore, $Q_{i} = Q = R^{c} \sin{\psi} \sim R \sin{\psi}$.
In addition, we expand the velocity field $\bm{v}^{f}_{i}(t)$ in \Eqsref{equ: velocity field} and \Braref{equ: force field acting on the fluid} up to second order in $(\zeta \bm{T})$ to obtain analytically tractable model equations.

Let us consider the correspondence between our simulation and
 the existing experiments\cite{suematsu2014quantitative, boniface2019self}.
First, we rewrite the reaction-diffusion equation \Eqref{equ:reaction-diffusion eq.} and the equation of motion \Eqref{equ:equation of motion} in non-dimensional forms by using the units of length, time and energy, $L_{0}, T_{0}$ and $E_{0}$, where these unit quantities are estimated for the experimental situation\cite{suematsu2014quantitative, boniface2019self} as $L_{0} \equiv R \sim 10^{-3}$ [m], $T_{0} \equiv \tau = m/\zeta$ [s], and $E_{0} \equiv 2 \pi \gamma_{0} R^{2} \sim 10^{-7}$ [J], respectively.
Here, $\tau$ is the relaxation time for the particle motion and can be estimated as $\tau \sim 1$ [s] for $m \sim 10^{-5}$ [kg].

The dimensionless parameters for the capillary interaction used in the simulations are
 $\psi \sim 0.1$ [rad] and $\text{Bo} \equiv (q R)^{2} \sim 0.1$ (determines the capillary length $q^{-1}$), the latter being related to the lateral capillary force [See \Eqref{equ:lateral capillary force}].
 Here, Bo represents the Bond number which is the ratio of the buoyancy force to the surface tension between the particle and the liquid.

On the other hand, the dimensionless quantities in reaction-diffusion equation for the surfactant [See \Eqref{equ:reaction-diffusion eq.}] are defined as follows;
\begin{align}
A^{*} = A T_{0}, \ 
\kappa^{*} = \kappa T_{0}, \ 
D^{*} = D \frac{T_{0}}{L_{0}^{2}},
\end{align}
where $D^{*}$ corresponds to the dimensionless diffusion length $\lambda = \sqrt{D/\kappa}$.
Furthermore, we define the following dimensionless quantities for the equation of motion for the particles [See \Eqref{equ:equation of motion}]
\begin{align}
\text{Pe} = \frac{vR}{D}, \ 
\text{Ma} = \frac{\alpha A R}{\mu D^{2}},
\label{equ: Pe and Ma}
\end{align}
where Pe is the P\'{e}clet number and is defined as the nondimensional self-propelling speed of the particles\cite{michelin2013spontaneous, boniface2019self}, and Ma is the solute Marangoni number, a dimensionless quantity defined by the ratio of the driving force of the concentration gradient to the viscous friction force.
Furthermore, in the direct particle-particle interaction potential, we define $\gamma_{0}^{*} = E_{0} T_{0} \sin^{2}{\psi}/(L_{0}^{2} \zeta)$, where $\gamma_{0}^{*}$ corresponds to the ratio of the lateral capillary force to the viscous friction force.

\section{Results and Discussion}
\label{sec:Simulation results}

\subsection{Single-particle system}
\begin{figure*}[htbp]
  \centering
  \includegraphics[clip, width=16.0cm]{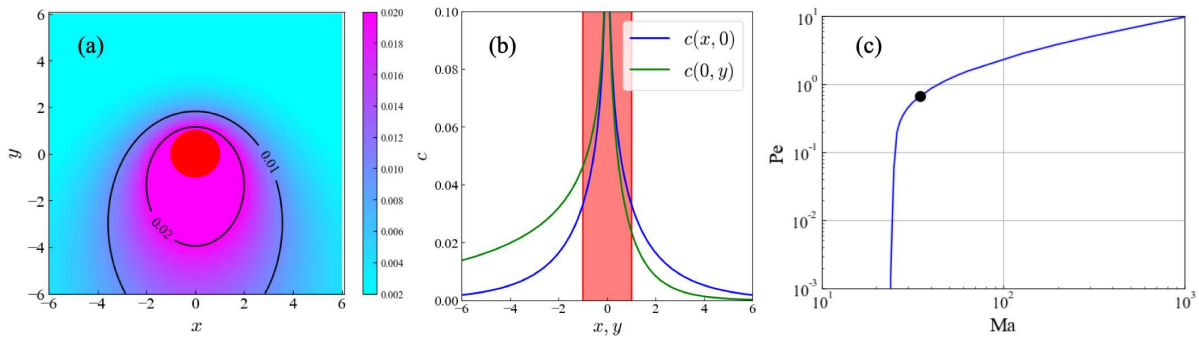}
  \caption{
   Self-driven motion for the single-particle system. (a) Color display of the concentration field of surfactant around the particle, where the particle is moving in the positive direction of the $y$-axis, (b) its cross-sections along $x$- and $y$-axes, and (c) swimming speed (Pe) as a function of Marangoni number.
   The parameters are set as $D^{*} = 0.15$ and $\kappa^{*} = 0.01$.
   The red circle represents the particle in (a), and the red region represents the inside of the particle in (b).
   The data shown in (a) and (b) and the dot in (c) correspond to the case with $\bm{v}^{f}=(0.0, -0.1)$.
  }
  \label{fig:delta}
\end{figure*}

First, we discuss the behavior of single-particle systems.
\Figureref{fig:delta} denotes (a) the color display of the concentration field of surfactant around a particle, (b) its cross-sections along $x$- and $y$-axes, and (c) swimming speed (Pe) as a function of Marangoni number.
The data shown in (a) and (b) are obtained using \Eqref{equ:concentration field for delta function}, and the curve shown in (c) is calculated using \Eqref{equ:equation of motion}.
Experimentally, the Marangoni number is controlled by changing the viscosity, for example by adding glycerin to water\cite{nagayama2004theoretical, boniface2019self}.

The non-dimensional equation of motion for a single particle is represented as
\begin{align}
\frac{d}{dt} v = - v + F^{\text{M-self}}(v),
\label{equ:eq. of motion for one particle}
\end{align}
where $v$ is the swimming speed of the particle.
Let us consider the range of stability of the steady state solution $v_{0}$ of \Eqref{equ:eq. of motion for one particle}.
To show such a condition, we expand $F^{\text{M-self}}(v)$ around $v = 0$ as
\begin{align}
F^{\text{M-self}}(v) =
 C^{(1)} v
+ C^{(3)} v^3 + \cdots,
\end{align}
where
\begin{align}
C^{(n)} = \frac{1}{n!} \left. \frac{\partial^{n}}{\partial v^{n}}F^{\text{M-self}}(v) \right|_{v = 0} \ \ \ n = 1, 2, ... \  .
\end{align}
Here, $F^{\text{M-self}}(0)=0$ and $C^{(n)} = 0$ for $n = 2,4,6,...$, because $F_{\text{M-self}}(v)$ is an odd function of $v$ due to the symmetry of the system.
Therefore, we obtain
\begin{align}
\frac{dv}{dt} &=
\left( - 1 + C^{(1)} \right) v
+ C^{(3)} v^{3} + \cdots,
\label{equ:EOM-third}
\end{align}
with which we can clarify the condition for stably self-propelling motion as
\begin{align}
	- 1 + C^{(1)} &> 0
	,
	\label{equ:condition_v}
	\\
	C^{(3)} &< 0
	.
	\label{equ:condition_v3}
\end{align}
Here,
\begin{align}
	C^{(1)} &= \frac{\text{Ma}}{12 \pi} K_{0}\left( \sqrt{\frac{\kappa^{*}}{D^{*}}} \right),
	\label{equ: coef_1}
	\\
	C^{(3)} &= \frac{ \text{Ma} }{ 384 \pi (D^{*})^2 }
	\left[
	K_{0}\left( \sqrt{ \frac{ \kappa^{*} }{ D^{*} } } \right)
	- 4 \sqrt{ \frac{ D^{*} }{ \kappa^{*} } }
	K_{1} \left( \sqrt{ \frac{ \kappa^{*} }{ D^{*} } } \right)
	\right].
\end{align}
We can confirm that the condition \Eqref{equ:condition_v3} is always satisfied because the parameters $D^{*}$ and $\kappa^{*}$ are positive and $K_{0}(x) - 4 K_{1}(x)/x < 0$ for $x > 0$.
When this condition is satisfied, the steady-state velocity obtained with the expansion shown in \Eqref{equ:EOM-third} up to the third order of the velocity $v$ is given by
\begin{align}
v_{0} = \sqrt{\frac{C^{(1)} - 1}{-C^{(3)}}}.
\label{equ:steady-state solution}
\end{align}
This result means that there is a threshold value for Ma where a single particle starts to move spontaneously (See \Figref{fig:delta}(c)).
Similarly to this result, the experimental data shows a positive correlation between Pe and Ma, and the self-propelling speed has a threshold value for Ma\cite{nagayama2004theoretical}.
Thus, our model can reproduce the onset of the self-propelling motion induced by a spontaneous symmetry breaking of the concentration field.

The self-driven velocity of each particle is given by \Eqref{equ:steady-state solution}, which means that the threshold is determined by $C^{(1)} = 1$.
Substituting \Eqsref{equ:condition_v} and \Braref{equ: coef_1} into this condition, the expression of the threshold value of Ma in the self-driven motion is obtained.
In this expression of the threshold value, $D^{*}$ and $\kappa^{*}$ play opposite roles, i.e. larger $D^{*}$ decreases the threshold value of Ma while larger $\kappa^{*}$ increases the threshold value of Ma.
These tendencies can be understood as follows.
When $D^{*}$ is large, the concentration distribution of surfactant spreads over a wide area in an asymmetric manner, resulting in an promoted mobility.
On the other hand, when $\kappa^{*}$ is large, the concentration distribution of surfactant decays faster before it spreads over a distance, resulting in a decay of the concentration gradient and a decrease in mobility.
In our simulations, we chose the values of the parameters $D^{*}$ and $\kappa^{*}$ based on the validation shown in \Appref{sec:validation of the model}.

In \Figref{fig:delta}(c), the self-propelling motion of each particle occurs when Ma is larger than a threshold value.
The definition of Ma in \Eqref{equ: Pe and Ma} means that a large Ma corresponds to a small viscosity $\mu$, a small diffusion constant $D$, a large emission rate of surfactant $A$, or a large reduction rate of the interfacial tension due to surfactant $\alpha$.
All these properties tend to enhance the asymmetric distribution of the surfactant around the particle, which leads to a large self-driving force.
When the self-propelling force exceeds the other viscous drag force induced by the Marangoni flow, the condition \Eqref{equ:condition_v} is satisfied and the particle starts to move spontaneously.

\subsection{Two-particle system}
\begin{figure}[htbp]
  \centering
  \includegraphics[clip, width=8.0cm]{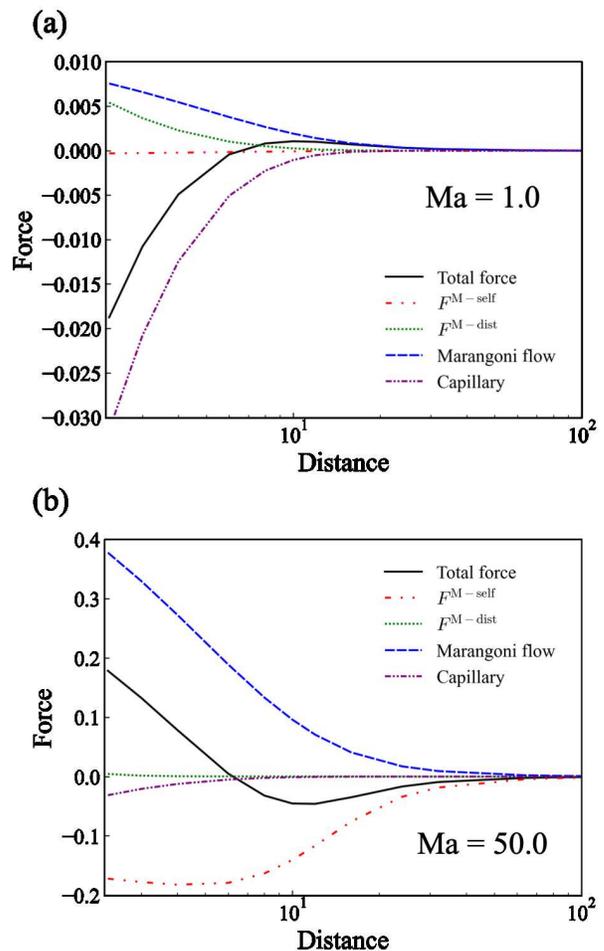}
  \caption{
  Forces acting between two particles for $D^{*} = 0.15, \kappa^{*} = 0.01, \gamma_{0}^{*} = 0.1$, and (a) Ma $= 1.0$ and (b) Ma $= 50.0$.
  ``Marangoni flow'' and ``Capillary'' denote the repulsive viscous force due to the Marangoni flow and the attractive force caused by capillary interaction, respectively.
  }
  \label{fig:force_dis}
\end{figure}

Before discussing the dynamical behavior of many particle systems, we investigate the elementary components of the interaction between two particles as functions of their separation distance.
\Figureref{fig:force_dis} shows the individual components of the force acting between two particles.
As was shown in \Figref{fig:delta}(c), the self-driven motion does not occur in \Figref{fig:force_dis}(a) ($\text{Ma}=1.0$) but occurs in \Figref{fig:force_dis}(b) ($\text{Ma} = 50.0$).
\Figureref{fig:force_dis} shows that there is a characteristic inter-particle distance where the attraction and repulsion are switched.
In \Figref{fig:force_dis} (a), the self-propelling force is small due to the small Ma as was shown in \Figref{fig:delta}(c).
For this reason, the main force acting between the two particles is the capillary force caused by the depression of the water surface.
As the interparticle distance increases, the capillary force decays rapidly, while the Marangoni flow decays slowly, leading to the dominance of the repulsive interaction induced by the Marangoni flow.
A competition between these two tendencies defines a threshold distance, inside and outside of which the interparticle interaction changes its nature from attractive to repulsive ones.

In the case of large Ma \Figref{fig:force_dis}(b), both self-propelling force and Marangoni flow are large (As Ma is proportional to $\alpha$, a large Ma means that the change in the interfacial tension caused by the surfactant is large.).
In this case, the behavior at short distances is dominated by the interplay between the self-propelling force and the viscous drag force due to the Marangoni flow.
In the initial rest state, the direction of the initial particle motion is determined by the self-propelling force induced by the high surfactant concentration in the interparticle region, which reduces the surface tension in that region.
This inhomogeneity in the surface tension generates an outward force due to the higher surface tension in the outside region, leading to an initial repulsive force.
Then, the Marangoni flow enhances this repulsive interaction.
On the other hand, at large interparticle distance, the capillary interaction and the self-propelling force $F^{\text{M-self}}$ determine the direction of the particle motion.
In the initial rest state, even though the capillary force is weak, it induces an attractive interaction, which leads to the particle motion toward the attractive direction.
This initial motion induces the self-driven motion in the attractive direction because the self-propelling force is insensitive to the separation between the particles.
Therefore, when the distance between the two particles is large, the self-propelling force acts as the attractive force.

\begin{figure*}[htbp]
  \centering
  \includegraphics[clip, width=16.0cm]{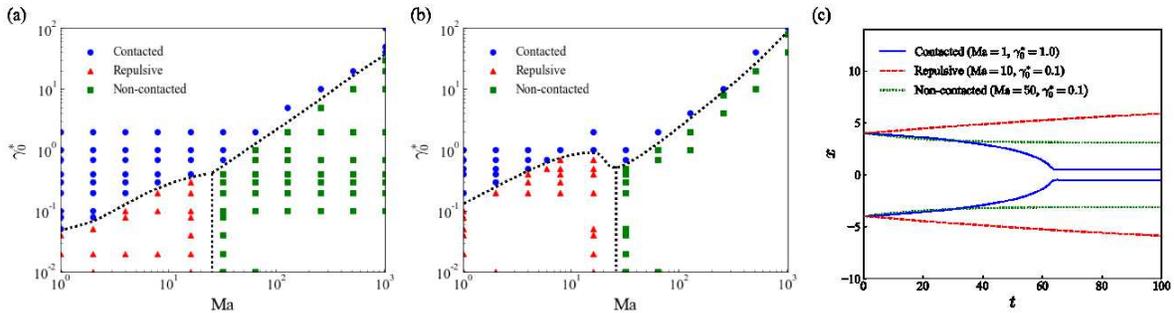}
  \caption{
  Phase diagrams and the typical trajectories of the two-particle systems for the relative motion of two particles for $D^{*} = 0.15$ and $\kappa^{*} = 0.01$. Phase diagrams are obtained for (a) $r_{1 2}(0) = 3.0$ and (b) $r_{1 2}(0) = 8.0$, where $r_{1 2}(0)$ denotes the inter-particle distance at $t = 0$ and the phase boundaries are the guide to the eyes.
  Figure (c) shows trajectories of the two particles starting from $r_{1 2}(0) = 8.0$ for various values of Ma and $\gamma_{0}^{*}$.
  }
  \label{fig:phase_diag_two}
\end{figure*}

\Figureref{fig:phase_diag_two} shows phase diagrams and trajectories of individual cases for the two-particle system.
These phase diagrams are made under the initial conditions (a) $r_{1 2}(0)=3.0$ (inside of the characteristic inter-particle distance) and (b) $r_{1 2}(0)=8.0$ (outside of the characteristic distance) (See \Figref{fig:force_dis}).
The states of two-particle system are classified into three motions; contacted, repulsive, and non-contacted motions, respectively.
The trajectories of each motion are shown in \Figref{fig:phase_diag_two}(c).
Additionally, the boundary line between the regions of repulsive and non-contacted motions is determined by the threshold in the motion of the single particle system shown for \Figref{fig:delta}(c).

\subsection{Many-particle system}
\begin{figure}[htbp]
  \centering
  \includegraphics[clip, width=8.0cm]{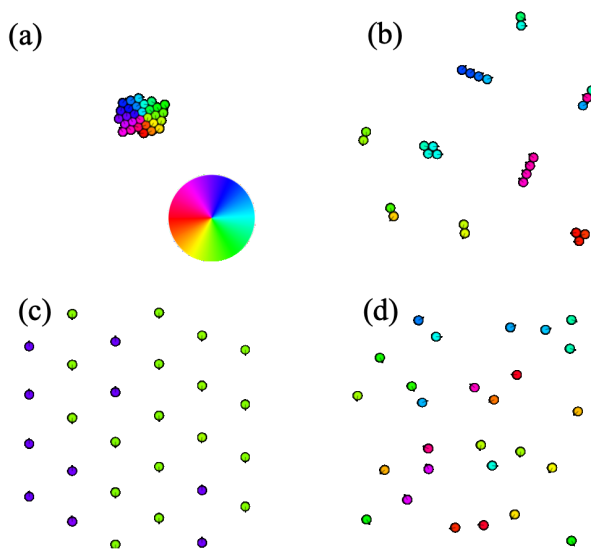}
  \caption{
  Four typical collective motions;
  (a) crystalline, (b) small chains and clusters, (c) spreading and (d) liquid-like states, respectively.
  The directions of velocities of the particles are shown by arrows and colors defined in the color legend in (a).
  }
  \label{fig:collective_motion}
\end{figure}

Let us discuss the behavior of collective motion of the particles by showing simulation results, such as phase diagrams and order parameters.
In \Figref{fig:collective_motion}, we show 4 typical collective behaviors; i.e. crystalline, chain, spreading and liquid-like states. (See the Supplemental Material \cite{supmovie} for movies of these behaviors.)
These collective behaviors correspond to experimentally observed states.
For example, the crystalline (\Figref{fig:collective_motion}(a)) and the chain states (\Figref{fig:collective_motion}(b)) reproduce the collective dynamics of self-propelled droplets\cite{tanaka2017dynamic}, and the spreading state (\Figref{fig:collective_motion}(c)) corresponds to the dynamic self-organization of the camphor disks\cite{soh2008dynamic}.

In order to identify the phase boundaries of each state shown in \Figref{fig:collective_motion}, we introduce several order parameters to characterize the collective behaviors.
The crystalline state can be characterized by the 6-fold bond-orientational order parameter\cite{bialke2015active} defined by
\begin{align}
\psi_{6} &= \frac{1}{N} \sum_{j = 1}^{N} |\psi_{6}^{j}|, \\
\psi_{6}^{j} &= \frac{1}{Z_{j}} \sum_{k}^{Z_{j}} e^{i6\theta_{jk}},
\end{align}
where $Z_{j}$ is the coordination number of $j$-th particle obtained from a Voronoi construction for the particle configuration, and $\theta_{jk}$ is the angle between a reference axis and the direction of the bond between $j$-th particle and its $k$-th neighbor. $\psi_{6} = 1$ means perfect hexagonal ordering, whereas completely disordered structures give $\psi_{6} = 0$.

As the above bond-orientational order parameter $\psi_{6}$ gives a large value not only for crystalline structure but also a straight string-like structures, $\psi_{6}$ cannot distinguish chain state from crystalline state.
Thus, we introduce an additional orientational order parameter defined by
\begin{align}
	\phi &= \frac{1}{M} \sum_{i} \phi^{i}, \\
	\phi^{i} &= \frac{1}{{}_{ {}^{\mathcal{N}_{i}} } C_{2}} \sum_{(j, k) \in \mathcal{S}_{i}, j \neq k} \left[ \frac{2}{3} \left( \frac{1}{2} - \frac{\bm{r}_{ij} \cdot \bm{r}_{jk}}{r_{ij}r_{jk}} \right) \right], 
\end{align}
where $M$ represents the total number of particles that are in contact with two or more other particles, $\mathcal{N}_{i}$ is the number of particles in contact with the $i$-th particle, ${}_{{}^{\mathcal{N}_{i}}}C_{2}$ is the binomial coefficient, and $\mathcal{S}_{i}$ denotes the search region for the other particles that contact the $i$-th particle, i.e. a circle with a radius of the order of the particle diameter centered at the center of mass of $i$-th particle.
The order parameter $\phi$ is defined for clusters composed of three or more particles, and this  parameter takes the value $\phi = 1$ in the case of the chain state, and $\phi = 0$ in the case of the crystalline state.
Here, the order parameter $\phi^{i}$ is defined for $i$-th particle, where $\phi^{i} = 1$ in the case that the relative positions of neighboring particles of the $i$-th particle align in a straight line.

\begin{figure*}[htbp]
  \centering
  \includegraphics[clip, width=16.0cm]{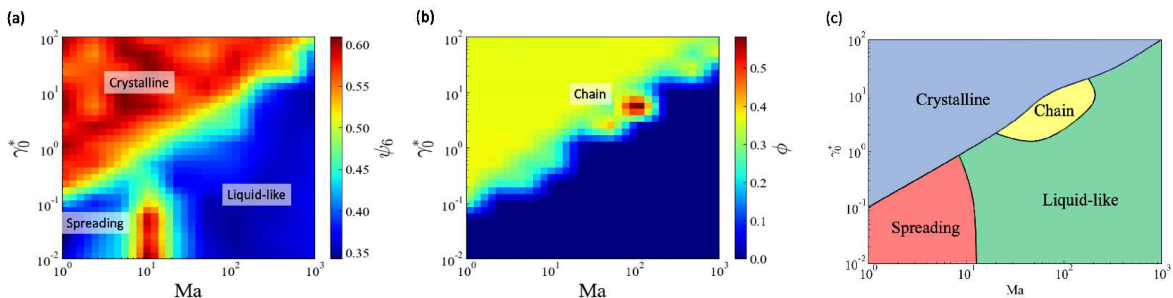}
  \caption{
  Stable regions of individual states on the Ma-$\gamma_{0}^{*}$ plane for $N = 30$ systems.
  The stable regions are determined with use of the order parameters,
  (a) bond-orientational order parameter $\psi_{6}$ and (b) orientational order parameter $\phi$.
  In this analysis, we averaged over 50 runs starting from different initial configurations.
  (c) From the order parameters, we can draw phase diagram for many-particle systems, where the phase boundaries are guide to the eyes.
  }
  \label{fig:order_parameter}
\end{figure*}

\Figureref{fig:order_parameter} shows the phase diagram of parameter regions for each state determined using the order parameters introduced above.
Here, we take the Marangoni number Ma and the magnitude of the capillary force $\gamma_{0}^{*}$ as independent parameters.
In \Figref{fig:order_parameter}(a), the capillary interaction is dominant in the crystalline state.
On the other hand, when the capillary force is small, the states are classified according to the Marangoni number into spreading and liquid-like states.
The boundary between the spreading state and the liquid-like state locates at the threshold value of Ma for the self-driven motion found in \Figref{fig:delta}(c).
This means that the spreading state is dominated by the Marangoni flow only, and the liquid-like states are driven by the self-driven motion.
In \Figref{fig:order_parameter}(b), the chain state is realized in a limited domain of Ma $\sim 100.0$ and $\gamma_{0}^{*} \sim 5.0$.
This means that the chain state is caused by a balance of all physical factors introduced in our model; i.e. hydrodynamic interaction, capillary interaction, and Marangoni effect.

We show the phase diagram for many-particle system in \Figref{fig:order_parameter}(c).
The boundaries of this diagram are guide to the eyes that are determined based on the results of the order parameter calculations shown in \Figsref{fig:order_parameter} (a) and (b).
The region of contacted, repulsive, and non-contacted motions for two-particle system in \Figref{fig:phase_diag_two} correspond to the crystalline, diffusion, and liquid states in \Figref{fig:order_parameter}, respectively.
The region of the chain state is a newly added phase in the many-particle system.

In the region of large $\gamma_{0}^{*}$, the attractive capillary force due to the depression of the water surface causes the particles to coagulate into a crystalline state.
When $\gamma_{0}^{*}$ is small, the attractive capillary fore is small, and the Marangoni effect is dominant.
When Ma is large, the self-propelling force becomes large, and the particles start to move freely and realizes the liquid state. In this case, the Marangoni flow acts as a repulsive interparticle force, which prevents the particles to coalesce at short distance.
On the other hand, when Ma and $\gamma_{0}^{*}$ are small, particles are repelling with each other to form the spreading state.
The chain state occurs when both Ma and $\gamma_{0}^{*}$ are large.
Since $\gamma_{0}^{*}$ is large, there is a large attraction between the particles due to the depression of the water surface. This causes the particles to form clusters. Since Ma is also large, however, the particles try to keep moving.  Such a competition results in a chain state where the two mechanisms are well-balanced, i.e. the particles try to move spontaneously due to the Marangoni effect while the capillary interaction keeps the clusters.

%

\section{Conclusion}
\label{sec:conclusion}
We have developed a minimal model to describe chemically active particles at an air-liquid interface.
In the modeling, the interactions between particles are decomposed into hydrodynamic interactions, capillary interactions, driving forces due to Marangoni effect, and Marangoni flow.

Analysis on the equation of motion for the single-particle system showed the existence of a threshold for the self-driven motion when the viscosity of liquid is changed. (i.e. Ma is increased.)
This finding was confirmed by numerical calculations, where the model parameters are chosen so that the model corresponds to the experimental situation\cite{nagayama2004theoretical, boniface2019self}.
These results show that our simple model can capture the essential properties of the experimental systems.

For the two-particle systems, we discussed the behavior of the inter-particle interaction as a function of the separation distance, which shows a switching from repulsive to attractive at a certain threshold distance.
Such a separation distance dependence of the inter-particle interaction is in good agreement with the previous simulations reported by Soh \textit{et al.}\cite{soh2008dynamic} for the case without the capillary interaction.
Based on these results, we can construct the phase diagrams for these  two-particle systems, with which we can understand the elementary behavior of many-particle system.

Decomposing the inter-particle interaction into contributions from different physical elements listed in \Figureref{fig:intro_model}, we find that the concentration field and the repulsion by Marangoni flow give the most dominant contribution.

Simulations on many-particle systems reproduced the collective behaviors found in the existing experiments, from which we can identify the major contribution for each state, i.e. the spreading state is caused by Marangoni flow as a major physical factor, the crystalline state by capillary interaction, the liquid-like state by self-propelling force, and the chain state by all physical elements included in our model. 
Our model, which includes the effect of the hydrodynamic interactions, can reproduce both crystalline and chain states\cite{tanaka2017dynamic} and the self-organization of particles with regular intervals\cite{soh2008dynamic}.
Furthermore we could identify the stable region for each state expressed by two parameters, i.e. Marangoni number Ma and the amplitude of capillary interaction $\gamma_{0}^{*}$.

With a further extension and an improvement of our model by introducing time-dependent model parameters, we will try to explain time-dependent non-steady collective behaviors.
By using our model, one can explore the behavior of many-particle system, with which we hope to observe typical dynamic phenomena such as motility induced phase separation in chemically active particles.
Applying our model to phenomena with much larger length and time scales, such as scaling behaviors, will be another interesting extension of the current study, where a further coarse-graining and multiscale treatment would be necessary.

\begin{acknowledgments}
The present work is partially supported by the Grant-in-Aid for Scientific Research from The Ministry of Education, Culture, Sports, Science and Technology of Japan (Grant No. 19H01858). 
\end{acknowledgments}

\appendix

\section{The derivation of the steady state concentration field of surfactant}
\label{sec: The derivation of Green function for the concentration field of surfactant}
We consider the following reaction-diffusion equation for surfactant density $\Gamma(\bm{r}, t)$
\begin{align}
\frac{\partial \Gamma}{\partial t}
+
\bm{\nabla} \cdot \left( \Gamma \bm{v}_{f} \right)
=
D \bm{\nabla}^{2} \Gamma
- \kappa \left( \Gamma - c_{\infty} \right)
+ A \delta(\bm{r})
,
\label{equ:reaction-diffusion delta}
\end{align}
where $\delta(\bm{r})$ is Dirac's delta function.
We solve this equation under the quasi-static approximation where the velocity field $\bm{v}_{f}$ is assumed to be in a steady state.
Then, the concentration field of the surfactant at time $t$ generated at a previous time $t'$ can be expressed in terms of the Green's function.
This Green's function $\Gamma_{t'}(\bm{r}, t - t')$ is obtained by solving
\begin{align}
\frac{\partial}{\partial t} \Gamma_{t'}
+
\bm{v}_{f} \cdot \bm{\nabla}  \Gamma_{t'}
=
D \bm{\nabla}^{2} \Gamma_{t'}
- \kappa \Gamma_{t'}
+ A \delta(\bm{r}) \delta(t - t'),
\end{align}
where the concentration field for steady state of \Eqref{equ:reaction-diffusion delta} is given by
\begin{align}
\Gamma(\bm{r}) = \int_{-\infty}^{t} dt' \Gamma_{t'}	(\bm{r}, t - t')
.
\label{equ:green function of gamma}
\end{align}
We define the Fourier transform for position $\bm{r}$ as
\begin{align}
	\widetilde{G}_{k} (\bm{k}, t)
	&=
	\frac{1}{2 \pi}
	\int_{-\infty}^{\infty} \int_{-\infty}^{\infty} 
	G(\bm{r}, t)
	e^{- i \bm{k} \cdot \bm{r}}
	dx dy
	,
	\label{equ:FT_k}
	\\
	G(\bm{r}, t)
	&=
	\frac{1}{2 \pi}
	\int_{-\infty}^{\infty} \int_{-\infty}^{\infty} 
	\widetilde{G}_{k} (\bm{k}, t)
	e^{i \bm{k} \cdot \bm{r}}
	dk_{x} dk_{y}
	,
	\label{equ:inv_FT_k}
\end{align}
and for time $t$ as
\begin{align}
	\widetilde{G}_{\omega}(\bm{r}, \omega)
	&=
	\frac{1}{\sqrt{2 \pi}}
	\int_{- \infty}^{\infty} G(\bm{r}, t)
	e^{- i \omega t} dt
	,
	\label{equ:FT_w}
	\\
	G(\bm{r}, t)
	&=
	\frac{1}{\sqrt{2 \pi}}
	\int_{- \infty}^{\infty}
	\widetilde{G}_{\omega}(\bm{r}, \omega)
	e^{ i \omega t} dt
	.
	\label{equ:inv_FT_w}
\end{align}
Using these Fourier transforms, we can obtain the Green's function as
\begin{align}
\Gamma_{t'}(\bm{r}, s)
&=
\frac{A}{4 \pi D s}
\exp{\left[
-\left( \kappa s
+\frac{\left( \bm{r} - \bm{v}_{f}s \right)^{2} }{4 D s}
\right)
\right]}
.
\label{equ: gamma t}
\end{align}
Therefore, the following expression for the Green's function is obtained from \Eqsref{equ:green function of gamma} and \Braref{equ: gamma t};
\begin{align}
\Gamma(\bm{r})
&=
\frac{A}{2 \pi D} \exp{\left[ \frac{\bm{r} \cdot \bm{v}_{f}}{2D} \right]}
K_{0}\left(2 r \sqrt{\frac{\kappa}{4D} + \left( \frac{\bm{v}_{f}}{4D} \right)^{2}} \right)
,
\end{align}
where $K_{0}(x)$ is the $0$-th order modified Bessel function of the second kind.
Using the above Green's function, we can express the steady-state concentration field for the source with any shape $f(\bm{r})$ as
\begin{align}
c(\bm{r})
&=
\int d\bm{r}' f(\bm{r}') \Gamma(\bm{r} - \bm{r}')
.
\end{align}
%

\section{Derivation of the driving force caused by Marangoni effect}
\label{sec: The driving force caused by Marangoni effect}
The driving force due to Marangoni effect $\bm{F}^{\text{M}}_{i}$ is divided into two parts as $\bm{F}^{\text{M}}_{i} = \bm{F}^{\text{M-self}}_{i} + \bm{F}^{\text{M-dist}}_{i}$, where $\bm{F}^{\text{M-self}}_{i}$ is the self-propelling force and $\bm{F}^{\text{M-dist}}_{i}$ is the interaction force due to the surfactant distribution emitted by the neighboring particles.
In addition, we introduce driving force by Marangoni flow caused by the change of the surface tension of air-liquid interface\cite{dominguez2016effective}.

\subsection{Self-propelling force}
Let us consider self-propelling force caused by the Marangoni effects.
The local surface tension of the air-liquid interface is assumed to be linearly dependent on the surfactant concentration as
\begin{equation}
\gamma(\bm{r}) = \gamma_{0} - \alpha c(\bm{r}),
\label{equ:the surface tension for dependence on surfactant concentration}
\end{equation}
where this surface tension is defined at the interface between the liquid and the air.
By integrating the force due to this surface tension along the three-phase contact line (hereafter denoted as $C$) around the particle, we can obtain the self-propelling force as
\begin{equation}
\bm{F}^{\text{M-self}}
=
\oint_{C} (\gamma_{0} - \alpha c(\bm{r})) \widehat{\bm{r}} \ dl
=
- \alpha \oint_{C}   c(\bm{r}) \widehat{\bm{r}} \ dl,
\label{equ:self-propelling force due to Marangoni effect}
\end{equation}
where $\hat{\bm{r}}$ denotes unit normal vector from the center of mass of the particle to the point on the contact line.
This integration leads to the expression of the driving force as
\begin{align}
\bm{F}^{\text{M-self}}_{i}
= - \frac{\alpha A R}{D}
I_{1} \left( \frac{v^{f}_{i} R}{2D} \right)
K_{0} \left( W_{i} R \right)
\hat{\bm{v}}^{f}_{i}
,
\label{equ:self-propelling force (appendix)}
\end{align}
where $W_{i} = 2 \sqrt{\frac{\kappa}{4D} + \left( \frac{v^{f}_{i}}{4D} \right)^2}$, $R$ indicates the radius of the particle and $I_{1}(x)$ is the 1-st order modified Bessel function of the first kind.
Note that, when the $i$-th particle is moving with a velocity $\bm{v}^{p}_{i}$, we should replace $\bm{v}^{f}_{i}$ by $\bm{v}^{f}_{i} - \bm{v}^{p}_{i}$, where $\bm{v}^{f}_{i} = \bm{v}^{f}(\bm{r}^{p}_{i}(t), t)$.
 
 \subsection{The influence of the surrounding particles}
Let us consider the inter-particle force caused by the concentration field of the surfactant emitted by the other surrounding particles.
By using the steady state solution of the concentration field, we can easily describe the inter-particle forces as
\begin{align}
 \bm{F}^{\text{M-dist}}_{i}
 &=
 \sum_{j \neq i} \bm{F}^{\text{M-dist}}_{j \to i} ,
 \\
 \bm{F}^{\text{M-dist}}_{j \to i}
 &=
 - \alpha
\oint_{C_{i}}
c(\bm{r} + \bm{r}_{ij})
 \widehat{\bm{r}}
 \ dl
 \\
 &\sim
 - \alpha \oint_{C_{i}}
 R \bm{\nabla} c( \bm{r}_{ij} )
 \cdot \hat{\bm{r}} \hat{\bm{r}} dl
 ,
\end{align}
where $C_{i}$ is the contact line on the surface of the $i$-th particle.
We keep the leading order term in the expansion of the solution with respect to the ratio between the particle radius and the inter-particle distance $R/r_{ij}$.
Based on this leading order approximation, the inter-particle force is calculated as follows
\begin{align}
 \bm{F}^{\text{M-dist}}_{j \to i}
 &=
 - \frac{\alpha A R^{2}}{2D}
 \exp{\left[ \frac{\bm{v}^{f}_{i} \cdot \bm{r}_{ij}}{2D}\right]}
 \nonumber
 \\
 &\ 
 \times \left[
 K_{0}(W_{j} r_{ij}) \frac{\bm{v}^{f}_{i}}{2D}
 -
 W_{j} K_{1}(W_{j} r_{ij}) \hat{\bm{r}}_{ij}
 \right]
 ,
 \label{equ: concentration field interaction}
\end{align}
where $K_{0}(x)$ and $K_{1}(x)$ are the $0$-th and $1$-st order modified Bessel functions of the second kind, respectively.

\subsection{Repulsive interaction caused by the Marangoni flow}
\label{sec:Repulsive interaction caused by the Marangoni flow}
The Marangoni flow is induced by the unbalance of the Marangoni stress on the air-liquid interface, and is expressed as follows\cite{dominguez2016effective};
\begin{align}
\bm{u}(\bm{r}) = \int d\bm{r}' 2 \bm{G}_{0}(\bm{r} - \bm{r}') \cdot
\left[-  \alpha \bm{\nabla'}_{\parallel} \Phi(\bm{r}') \right],
\end{align}
where $\nabla_{\parallel} \equiv (\partial_{x}, \partial_{y})$, $2 \bm{G}_{0}(\bm{r})$ denotes Oseen tensor on the air-liquid interface, and $\Phi(\bm{r}) = \sum_{i} c(\bm{r} - \bm{r}^{p}_{i})$ represents the total concentration field at the location $\bm{r}$.
We can derive the Marangoni flow by solving the reaction-diffusion equation of surfactant \Eqref{equ:reaction-diffusion eq.} within the Stokes approximation as
\begin{align}
	u(\bm{r})
	=
	\sum_{i} \frac{A \alpha}{8 \mu D}
	f\left( \frac{|\bm{r} - \bm{r}_{i}^{p}|}{\lambda} \right)
	,
\end{align}
where $f(0) = 0$ and
\begin{align}
f(x) = \frac{L_{1}(x) + L_{-1}(x)}{2} - I_{1}(x)	 + \frac{1}{\pi}
\end{align}
for $x \neq 0$, and $L_{\nu}(x)$ is the modified Struve functions of order $\nu$.
We can confirm that the Marangoni flow at long distances decays as $1/r^2$ because $\bm{u}(\bm{r}) = \int d\bm{r}' \bm{G}_{0}(\bm{r} - \bm{r}') \cdot \bm{\nabla'}_{\parallel} \delta(\bm{r}') = - \bm{\nabla}_{\parallel} \bm{G}_{0}(\bm{r}) \propto 1/r^2$.

\section{Validation of the model}
\label{sec:validation of the model}
\begin{figure}[htbp]
  \centering
  \includegraphics[clip, width=8.0cm]{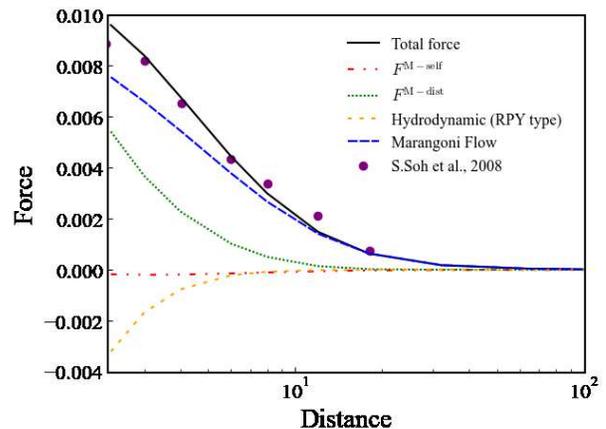}
  \caption{
  Distance dependences of the individual forces for the case with $D^{*}=0.15$, $\text{Ma} = 1.0$, $\kappa^{*} = 0.01$, and $\gamma_{0}^{*} = 0.0$.
  The black, pink, green, red, and blue lines represent total force acting on the particle, self-propelling force $F^{\text{M-self}}$, the inter-particle interaction force mediated by the surfactant concentration field $F^{\text{M-dist}}$, viscous friction force by the hydrodynamic interaction, and repulsive force by Marangoni flow, respectively.
  The dots show the data from the previous study by Soh \textit{et al.}\cite{soh2008dynamic} and the horizontal axis is in a logarithmic scale.
  }
  \label{fig:force dis gam0 detail}
\end{figure}

For a comparison between the present study and the previous studies\cite{soh2008dynamic}, we consider a case where the capillary interaction is neglected, i.e. $\gamma_{0}^{*} = 0.0$. (For the effect of capillary interaction, see \Figref{fig:force_dis}.)
\Figureref{fig:force dis gam0 detail} shows the dependences of individual forces on the particle distance for the case with $D^{*} = 0.15$.
The self-propelling force $F^{\text{M-self}}$ is almost zero for all distances, while the inter-particle force $F^{\text{M-dist}}$ mediated by the surfactant concentration field is large and attractive for short distances but is screened beyond the diffusion length $\lambda = \sqrt{D^{*}/\kappa^{*}} \sim 3.87$.
The repulsive force by the Marangoni flow gives the largest contribution to the total force, leading to the long-range repulsive interaction between particles.
%

\bibliographystyle{apsrev4-2}
\bibliography{manuscript}

\end{document}